\documentclass[pra,twocolumn,aps,floatfix,longbibliography,superscriptaddress]{revtex4-1}
\usepackage[caption=false]{subfig}
\usepackage{graphicx}
\usepackage{amsmath}
\usepackage{amssymb}
\usepackage{amsfonts}
\usepackage{enumerate}
\usepackage{color}
\usepackage{tikz}
\usepackage{ragged2e}
\usetikzlibrary{calc,decorations.markings}
\usepackage[colorlinks=true,linkcolor=blue,citecolor=blue,urlcolor=blue]{hyperref}
\usepackage{tipa}
\usepackage{physics}
\usepackage{soul}
\usepackage{float}
\usepackage{amsthm}

\newcommand{\x}{\mathsf{x}}

\usepackage{tikz,xcolor}
\definecolor{lime}{HTML}{A6CE39}
\DeclareRobustCommand{\orcidicon}{%
	\begin{tikzpicture}
	\draw[lime, fill=lime] (0,0) 
	circle [radius=0.16] 
	node[white] {{\fontfamily{qag}\selectfont \tiny ID}};
	\draw[white, fill=white] (-0.0625,0.095) 
	circle [radius=0.007];
	\end{tikzpicture}
	\hspace{-2mm}
}
\foreach \x in {A, ..., Z}{%
	\expandafter\xdef\csname orcid\x\endcsname{\noexpand\href{https://orcid.org/\csname orcidauthor\x\endcsname}{\noexpand\orcidicon}}
}
 
\begin{document}

\title{The gravitational quantum Otto refrigeration cycle}
\author{Nikos K. Kollas\orcidA{}}
\email{kollas@upatras.gr}
\date{\today}

\begin{abstract}
We take advantage of the gravitational redshift experienced by a photon propagating in curved spacetime in order to construct a quantum Otto refrigeration cycle. Deriving a lower bound for the relative temperature between the cold and hot reservoirs at which the engine operates, we provide examples of refrigeration cycles in the presence of a uniform gravitational field as well as in the case of a gravitational field produced by a non-rotating spherically symmetric charged body, and the one obtained from the vacuum solution of Einstein’s field equations in an expanding universe, as a function of the parameters which describe each spacetime.
\end{abstract}
\maketitle
\section{Introduction}
One of the most tested and successful theories of modern physics is that of general relativity which explains gravity in terms of the distortion of spacetime caused by a massive object and the apparent deviation of free-fall motion from a straight line. The solutions to Einstein’s field equations are in excellent agreement with observations and correctly describe various phenomena such as the precision of the perihelion of Mercury, gravitational lensing, the structure
of neutron stars, the accelerating expansion of the universe, gravitational waves \cite{Misner} as well as the existence of black holes, objects so dense that not even light can escape their gravitational pull.

During the seventies, several studies uncovered a fundamental connection between the physics of black holes and the laws of thermodynamics. In 1971 Hawking derived a second law for the surface area of an event horizon \cite{PhysRevLett.26.1344}, followed a couple of years later by Bekenstein’s formula for the entropy of a black hole \cite{PhysRevD.7.2333,PhysRevD.9.3292} and the four laws of black hole mechanics \cite{Bardeen1973}, culminating in the discoveries of Hawking and Unruh radiation \cite{Hawking1975,PhysRevD.14.870}. Hints towards a deeper relationship between gravity and thermodynamics were further uncovered by Jacobson who, in 1995, demonstrated that it is possible to derive the field equations from simple thermodynamic arguments \cite{PhysRevLett.75.1260}. 

Motivated by this connection the concept of a black hole heat engine was proposed \cite{10.1119/1.3633692,doi:10.1142/S0218271819500123,doi:10.1142/S0218271819500068} wherein work can, in principle, be produced through a Carnot cycle from two black holes of different masses by treating the mass M of the each hole as its internal energy, its surface gravity $\kappa$ as its temperature (through Hawking’s formula $T_H=\kappa/2\pi$ in natural units) and its horizon surface area $A$ as its entropy (through Bekenstein’s $S=A/4$ formula in natural units). Taking advantage of the AdS/CFT correspondence, Johnson suggested the notion of a holographic heat engine where the working medium comprises of a high temperature sector of the gauge theory to which the gravitational physics in anti-de Sitter space is dual \cite{Johnson_2014,Johnson_2020}. Alternate approaches involve the use of an effective heat bath observed by a two level system when it is accelerating through the vacuum of a scalar field for the construction of a so called Unruh quantum heat engine (with an associated Unruh temperature $T_U=a/2\pi$ where $a$ is the acceleration of the system in natural units) \cite{Gray2018,Arias_2018,Unruh:Otto:degen,Unruh:Otto:entangl,Unruh:Otto:entangl2,Unruh:Otto:reflect} and the use of confined quantum systems whose thermodynamic properties change with respect to gravitational degrees of freedom, as in the case of a quantum field in a cavity \cite{BRUSCHI2020126601} or a quantum particle in a gravitational well \cite{Santos2018}. In the latter approach \cite{Santos2018} the operation of an effective quantum Otto heat engine \cite{PhysRevE.61.4774,Kieu2006,Deffner:TMs} was described in which the adiabatic expansion and compression strokes of the cycle where constructed by controlling the strength of the gravitational field. Though in principle, it is always possible to perform such a control by appropriately accelerating the system the process can no longer be considered adiabatic since in this case the accelerated particle will experience vacuum thermal radiation from. Indeed this is precisely the idea behind the operation of the Unruh engine \cite{Arias_2018,Gray2018,Unruh:Otto:degen,Unruh:Otto:entangl,Unruh:Otto:entangl2,Unruh:Otto:reflect}.

In this letter we provide a proof of principle approach for a gravitational Otto engine which operates between two thermal reservoirs in different locations of a static gravitational field whose working medium comprises of photons, emitted and absorbed by two level systems, which experience gravitational redshift as they propagate from one position to the other. This approach, which permits a more general investigation of the effects of spacetime metrics on the efficiency and operation of heat engines, should be amenable to experimental implementation as quantum optical systems, which describe the interaction between light and matter, have already been experimentally demonstrated as promising platforms for quantum heat engines \cite{Deffner:TMs} and are also the basis for upcoming quantum technologies in space \cite{Kaltenbaek2021,Mohageg2022}.

We begin with the description of a quantum Otto refrigeration cycle which operates in the gravitational field of the Earth. The more general case of a static metric is treated later on. In the following we employ the natural system of units $c=\hbar=G=k_B=\varepsilon_0=1$ in which time, energy, mass, temperature and charge are all expressed in the same units as those of length.
\section{Refrigeration cycle in the gravitational field of the Earth}
\begin{figure*}
\centering
\includegraphics[width=\textwidth]{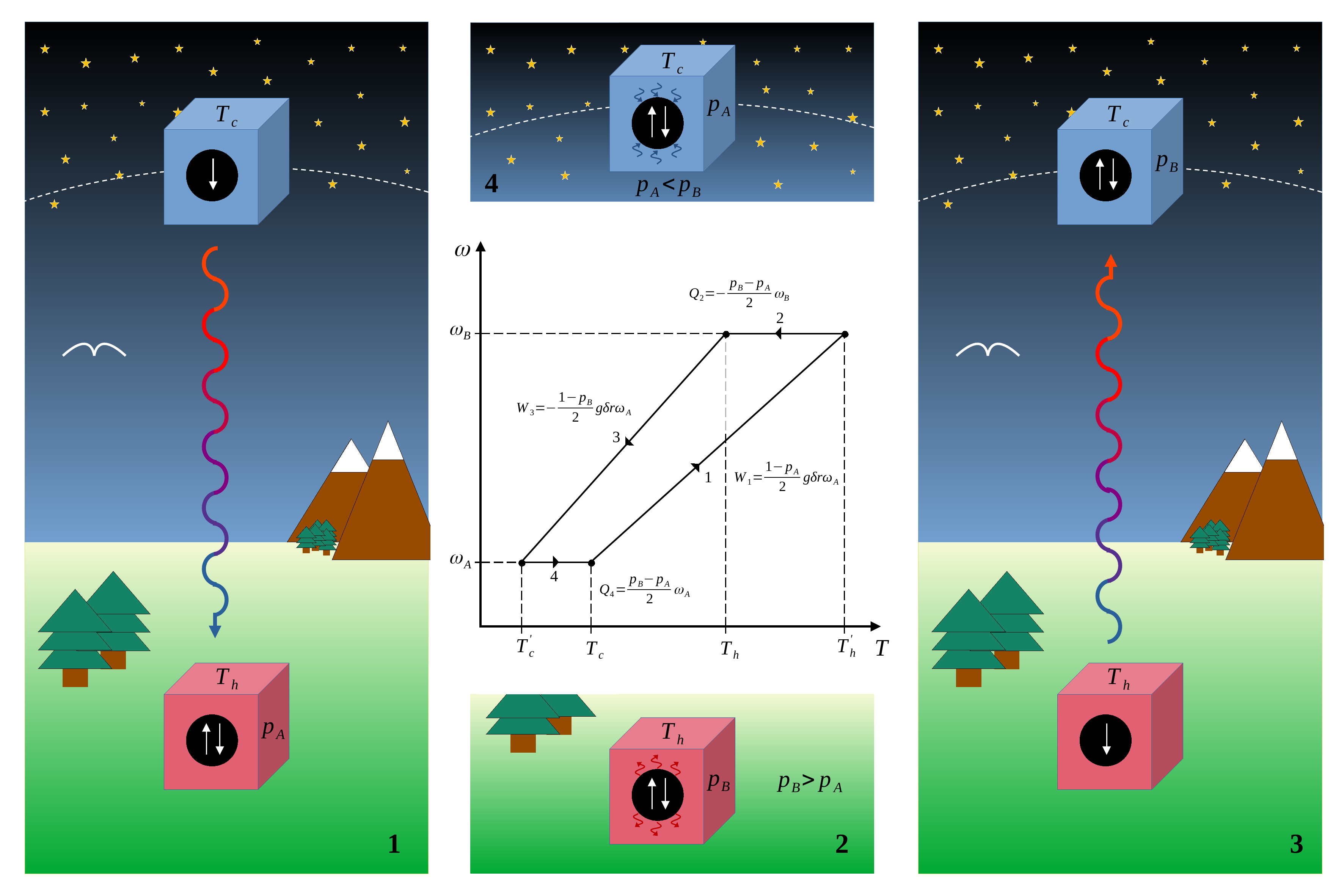}
\caption{The gravitational quantum Otto refrigeration cycle. {\textbf{1.}} A two level system aboard a satellite in a geostationary orbit emits a photon which, due to the gravitational field of the Earth, is blueshifted from $\omega_A$ to $\omega_B$ before being absorbed by a second qubit on the surface. {\textbf{2.}} The surface qubit thermalizes, thus increasing its purity form $p_A$ to $p_B$, by coming into isochoric contact with a thermal reservoir at temperature $T_h$. {\textbf{3.}} By transitioning to its ground state it then emits a photon towards the satellite which is redshifted back to $\omega_A$. {\textbf{4.}} By coming into isochoric contact with a thermal reservoir at temperature $T_c$ the orbiting qubit restores its temperature and purity thus completing the cycle. {\bf{Center:}} Otto refrigeration cycle in a temperature-energy diagram with closed expressions for the amounts of work and heat exchanged in each stroke of the cycle.} 
\label{GrOtto}
\end{figure*}
Consider the following thermodynamic cycle carried out by two observes, Alice and Bob, depicted in Fig. (\ref{GrOtto}). A two level system (qubit) with energy gap $\omega_A$ between its levels and initial state
\begin{equation}\label{in-state}
	\rho_A = \frac{1-p_A}{2}|\omega_A\rangle\langle\omega_A|+\frac{1+p_A}{2}|0\rangle\langle 0|,
\end{equation}
with 
\begin{equation}\label{purity}
	p_A=\tanh\frac{\omega_A}{2T_c}
\end{equation} 
is in thermal equilibrium with Alice's laboratory at temperature $T_c$ orbiting inside a satellite in a geostationary orbit around the Earth. By transitioning into its ground state the qubit emits a photon which is then absorbed by a second two level system with energy $\omega_B$ in Bob's laboratory which is located on the surface. Assuming the two laboratories lie on the same radial direction and are separated by a vertical distance $\delta r$ from each other then, due to the redshift experienced by the photon by the gravitational field of the Earth \cite{rodriguez2023introduction}, it is necessary for Bob's qubit to have an energy gap equal to
\begin{equation}\label{shift}
	\omega_B = (1+g\delta r)\omega_A
\end{equation}
in order to absorb the emitted photon, where $g$ is the acceleration on the surface of the Earth. If the initial state of Bob's qubit is in the ground state $\ket{0}$ then its final state will be equal to
\begin{equation}\label{in-state}
	\rho_B = \frac{1-p_A}{2}|\omega_B\rangle\langle\omega_B|+\frac{1+p_A}{2}|0\rangle\langle 0|.
\end{equation}
Replacing $\omega_A$ with $\omega_B$ in Eq. (\ref{purity}) we find that the temperature of Bob's qubit is similarly shifted by the same amount and is now equal to 
\begin{equation}\label{Tolman}
T'_h=(1+g\delta r)T_c.
\end{equation} 
This is simply the \emph{Tollman temperature} of the laboratory of Alice as measured by Bob \cite{Rovelli_2011,Santiago_2019}.

For the next stage of the cycle Bob brings his qubit in thermal contact with a reservoir at temperature $T_h$ which is less than the Tollman temperature
\begin{equation}\label{condition}
T_h<(1+g\delta r)T_c.
\end{equation} 
By reaching thermal equilibrium the qubit's purity is changed to
\begin{equation}
	p_B=\tanh\frac{\omega_B}{2T_h}>p_A.
\end{equation}
The qubit then decays into its ground state and emits a photon towards Alice's laboratory whose frequency is red-shifted back to its initial value $\omega_A$. Similar reasoning to that which led to the derivation of Eq. (\ref{Tolman}) implies that the temperature of Alice's qubit in her frame of reference aboard the satellite is now equal to
\begin{equation}
T'_c=\frac{T_h}{1+g\delta r}<T_c.
\end{equation}
The cycle terminates by allowing Alice's qubit to exchange heat with the cold reservoir restoring its temperature $T_c$ and purity $p_A$ back to their initial values.

Alice and Bob have thus realized a \emph{Quantum Otto refrigeration engine} \cite{PhysRevE.61.4774,Kieu2006,Deffner:TMs} where the energy gained or lost by the photon due to the presence of the gravitational field comprises the adiabatic stages of the cycle and the heat exchanged with the reservoirs the isochoric stages. By simply tracking the changes in the energy of the photon and the qubits it is easy to see that during the first adiabatic stroke of the cycle 
\begin{equation}\label{Win}
	W_1 = \frac{1-p_A}{2}g\delta r\omega_A
\end{equation}
units of work are consumed on average by the gravitational field in order to raise the energy of the photon, while 
\begin{equation}\label{Qout}
	Q_2 = -\frac{p_B-p_A}{2}\omega_B
\end{equation}
units of heat are released into the hot reservoir during equilibriation of Bob's qubit. Similarly 
\begin{equation}\label{Wout}
	W_3 = -\frac{1-p_B}{2}g\delta r\omega_A
\end{equation}
units of work are done against the photon which loses energy during transmission between Bob's and Alice's laboratory, while
\begin{equation}\label{Qin}
	Q_4 = \frac{p_B-p_A}{2}\omega_A
\end{equation}
units of heat are drawn by Alice's qubit aboard the satellite. The total change in energy is equal to zero
\begin{equation}
	W_1+Q_2+W_3+Q_4=0
\end{equation}
as should be expected for a closed thermodynamic cycle.

A merit of performance of a refrigeration cycle is given by the \emph{coefficient of performance} (COP)
\begin{equation}\label{COP}
	COP = \frac{Q_4}{W_1+W_3}=\frac{1}{g\delta r}.
\end{equation}
From Eq. (\ref{condition})
\begin{equation}\label{bound}
	\frac{1}{g\delta r}<\frac{T_c}{T_h-T_c}.
\end{equation}
The right hand side of Eq. (\ref{bound}) is the COP of the Carnot refrigeration cycle. Solving for the ratio between temperatures we find that for a given temperature of the hot reservoir the temperature of the cold reservoir ranges between
\begin{equation}\label{range}
	\frac{T_h}{1+g\delta r}<T_c<T_h.
\end{equation}
For a geostationary orbit $\delta r\approx 3.5\times 10^7 m$. Since $g\approx 10^{-16}m^{-1}$ it follows that although gravitational refrigeration near the earth is in principle possible, the lower limit for the allowed range of temperatures of the cold reservoir to first order is $(1-10^{-9})T_h$ and is thus of a negligible magnitude.
\section{Refrigeration cycle in a static gravitational field}
In a static gravitational field with metric $g_{\mu\nu}$ Eq. (\ref{shift}) is replaced by \cite{Misner}
\begin{equation}\label{general-shift}
	\omega_B = \frac{\sqrt{\abs{g_{00}^A}}}{\sqrt{\abs{g_{00}^B}}}\omega_A
\end{equation}
where $g_{00}$ denotes the purely timelike component of the metric tensor and superscripts indicate its value at Alice's and Bob's local frames of reference. Substituting $g \delta r\to \frac{\sqrt{\abs{g_{00}^A}}}{\sqrt{\abs{g_{00}^B}}}-1$ in Eqs (\ref{Win}), (\ref{Wout}) and (\ref{COP})-(\ref{range}) we find that for a given temperature of the hot reservoir the allowable range in temperatures of the cold reservoir in this case is given by
\begin{equation}\label{general-range}
	\frac{\sqrt{\abs{g_{00}^B}}}{\sqrt{\abs{g_{00}^A}}}<\frac{T_c}{T_h}<1.
\end{equation}

We will now consider two examples of static spacetimes described by the \emph{Reissner–Nordström} and \emph{De Sitter} metrics.
\subsection{Reissner–Nordström spacetime}
\begin{figure}
\centering
\includegraphics[width=\columnwidth]{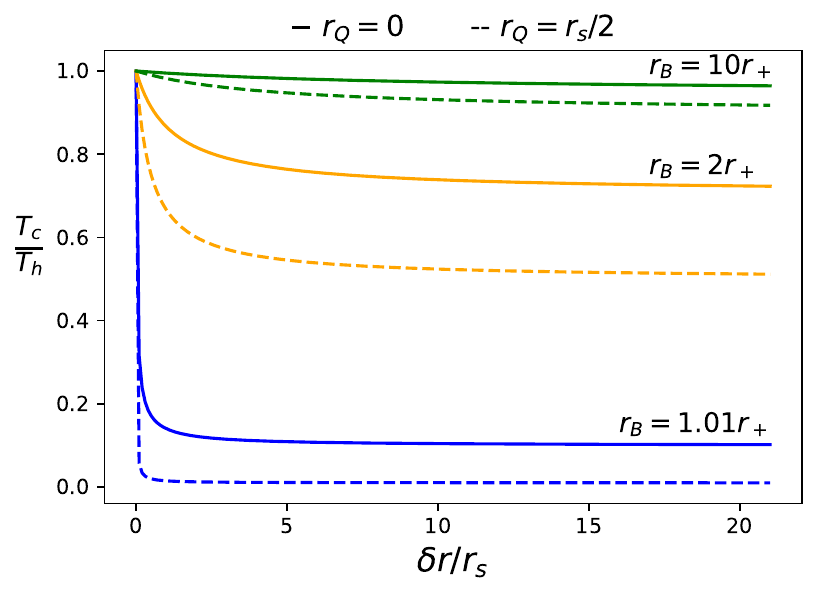}
\caption{Lower bounds for the ratio between the cold and hot temperatures above which a gravitational quantum Otto refrigeration cycle is possible in a Reissner-Nordström metric for different distances of Bob's frame of reference from the event horizon as a function of the radial separation $\delta r$ between Bob and Alice in Schwarzhild radius units.}
\label{Reissner}
\end{figure}
The metric corresponding to the gravitational field of a charged non-rotating spherically symmetric body of mass M is described by the Reissner–Nordström metric
\begin{equation}\label{metric}
ds^2 = -f(r)dt^2+\frac{dr^2}{f(r)}+r^2d\theta^2+r^2\sin^2\theta d\phi^2,
\end{equation}
with 
\begin{equation}
	g_{00}(r)=f(r)=1-\frac{r_s}{r}+\frac{r^2_Q}{r^2}
\end{equation}
where $r_s=2M$ is the \emph{Schwarzild radius} and $r^2_Q=\frac{Q^2}{4\pi}$ is a characteristic length associated with the body's charge. If Alice and Bob lie on the same radial direction at distances $r_A$ and $r_B$ outside the event horizon
\begin{equation}
	r_+=M+\sqrt{M^2-\frac{Q^2}{4\pi}},
\end{equation}
they can realize a gravitational refrigeration engine whose relative temperature is bounded by (assuming $r_A>r_B$)

\begin{equation}\label{general-range}
	\frac{\sqrt{1-\frac{r_s}{r_B}+\frac{r_Q^2}{r_B^2}}}{\sqrt{1-\frac{r_s}{r_A}+\frac{r_Q^2}{r_A^2}}}<\frac{T_c}{T_h}<1.
\end{equation}

In Fig. (\ref{Reissner}) we present the lower bound for different distances of Bob's frame of reference from the event horizon as a function of the radial separation $\delta r=r_A-r_B$. We observe that the range of temperatures at which the refrigerator can operate increases the closer Bob is to the horizon and the larger the separation between the laboratories. For a charged body, for which the radius of the event horizon decreases, this effect is enhanced even further.
\subsection{De Sitter spacetime}
\begin{figure}
\centering
\includegraphics[width=\columnwidth]{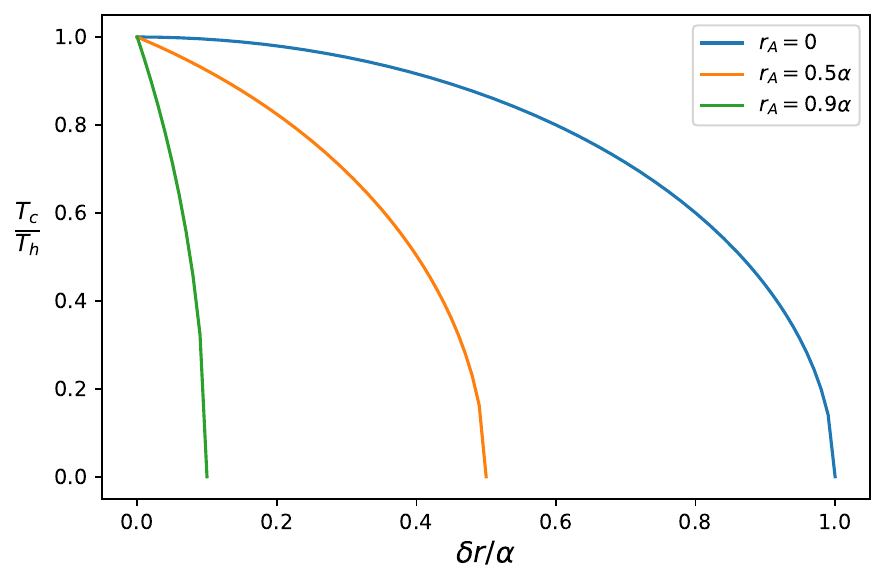}
\caption{Lower bounds for the ratio between the cold and hot temperatures above which a gravitational quantum Otto refrigeration cycle is possible in a De Sitter metric for different distances of Alices's frame of reference s a function of the radial separation $\delta r$ between Bob and Alice in cosmological horizon units $a$.}
\label{DeSitter}
\end{figure}

The metric in a De Sitter spacetime has the same form as Eq. (\ref{metric}) only this time
\begin{equation}
	f(r) = 1-\frac{r^2}{a^2}
\end{equation}
where $a$ is the cosmological horizon. The relative temperature of the gravitational refrigeration engine in this case is lower bounded by
\begin{equation}
	\frac{\sqrt{1-\frac{r_B^2}{a^2}}}{\sqrt{1-\frac{r_A^2}{a^2}}}<\frac{T_c}{T_h},
\end{equation}
note that in this case in order for $T_c/T_h<1$, $r_A<r_B$. In Fig. (\ref{DeSitter}) we present the lower bound for different distances of Alice's frame of reference from the event horizon as a function of the radial separation $\delta r=r_B-r_A$. We observe that the range increases the closed Bob gets to the cosmological horizon, in this case though small changes in separation result in large changes in the range of operational temperatures.

\section{Conclusions}
A first principles approach of a gravitational quantum Otto refrigeration cycle was given where the gravitational redshift experienced by photons propagating in a gravitational field was employed as a source of useful work for the engine. For any static spacetime it is easy to compute a lower bound for the temperature ratio between the cold and hot reservoirs. In general, as should be perhaps expected, the operation of the engine depends on the strength of the gravitational field. Although refrigeration is negligible near the Earth, it is strongly enhanced the closer the hot reservoir is located to the event horizon of a Reissner–Nordström spacetime. It would be of particular interest to inquire whether it is possible to reverse this process in order to extract a net amount of useful work from the gravitational field by operating an Otto heat engine. This kind of question is not new but rather dates back to the 19th century with Loschmidt's idea of extracting an infinite amount of energy by taking advantage of the temperature gradient in an atmospheric column, although the second law of thermodynamics suggests that the answer to this questions is negative, it is still open to this day \cite{Dreyer2000}.
\acknowledgments{The author wishes to thank Dimitris Moustos and Charis Anastopoulos for fruitful discussions. Part of this work was presented in the 14th international conference of Relativistic Quantum Information North.}
\bibliography{references}
\end{document}